\documentclass[%
 reprint,
nofootinbib,
 amsmath,amssymb,
 aps,
floatfix,
]{revtex4-2}
\usepackage{xcolor}

\usepackage{graphicx}
\usepackage{sidecap}
\usepackage{dcolumn}
\usepackage{bm}

\usepackage[utf8]{inputenc} 
\usepackage[T1]{fontenc}    
\usepackage[hidelinks]{hyperref}       
\usepackage{url}     
\usepackage{booktabs}       
\usepackage{siunitx}
\usepackage{amsfonts}       
\usepackage{nicefrac}       
\usepackage{microtype}      
\usepackage{lipsum}		
\usepackage{tikz}
\usepackage{mathrsfs} 
\usepackage{amssymb}
\usepackage{cleveref}
\usepackage{academicons}
\usepackage{enumerate} 
\definecolor{orcidlogocol}{HTML}{A6CE39}

\newcounter{Snumber}

\crefname{Sequation}{Eq.~(S\theSequation)}{Eqs.~(S\theSequation)}

\newcounter{Sfigure}

\newcounter{STable}

\crefname{Sfigure}{Fig.}{Figs.}
\crefname{SSCfigure}{Fig.}{Figs.}
\crefname{STable}{Tab.}{Tabs.}

\AtBeginDocument{%
    \crefname{equation}{Eq.}{Eqs.}%
    \crefname{chapter}{Ch.}{chapters}%
    \crefname{section}{Sect.}{Sects.}%
    \crefname{appendix}{appendix}{appendices}%
    \crefname{enumi}{item}{items}%
    \crefname{footnote}{footnote}{footnotes}%
    \crefname{figure}{Fig.}{Figs.}%
    \crefname{table}{Tab.}{Tabs.}%
    \crefname{theorem}{theorem}{theorems}%
    \crefname{lemma}{lemma}{lemmas}%
    \crefname{corollary}{corollary}{corollaries}%
    \crefname{proposition}{proposition}{propositions}%
    \crefname{definition}{definition}{definitions}%
    \crefname{result}{result}{results}%
    \crefname{example}{example}{examples}%
    \crefname{remark}{remark}{remarks}%
    \crefname{note}{note}{notes}%
}

\newcommand{\del}{\partial}

\makeatletter
\newcommand*{\transpose}{%
  {\mathpalette\@transpose{}}%
}
\newcommand*{\@transpose}[2]{%
  \raisebox{\depth}{$\m@th#1\intercal$}%
}
\makeatother

\newcommand*{\Det}{\operatorname{Det}}

\begin{document}

\preprint{APS/123-QED}

\title{
Curvature instability of an active gel growing on a wavy membrane}

\author{Kristiana Mihali}
\author{Dennis W{\"o}rthm{\"u}ller}%
\author{{Pierre Sens}}%
\email{pierre.sens@curie.fr}
\affiliation{ Institut Curie, Université PSL, Sorbonne Université, CNRS UMR168, Physique des Cellules et Cancer, 75005 Paris, France}


\date{\today}

\begin{abstract}
Cell shape changes are largely controlled by the actin cytoskeleton, a dynamic filament network beneath the plasma membrane. Several cell types can form extended free-standing protrusions not supported by an extracellular substrate or matrix, and regulated by proteins that modulate cytoskeletal dynamics in a way sensitive to the curvature of the cell membrane.
We develop a theoretical model for the mechanics of a free-standing viscous actin network growing on a corrugated membrane. The model couples the dynamics of the viscous active gel with membrane deformation and the recruitment of curvature-sensitive actin nucleators.
We show that an actin layer polymerising uniformly on the membrane always exerts a stabilising effect that reduces membrane deformation. However, curvature-sensitive actin nucleator proteins can render the membrane linearly unstable, depending on the interplay between membrane and actin dynamics, giving rise to spontaneous membrane deformation which could initiate extended free-standing cellular protrusion.
\end{abstract}

\maketitle

Cell shape change is essential for a wide range of cellular functions, including migration and morphogenesis. It relies on the mechanical stress generated by the cytoskeleton on the cell membrane \cite{lecuit2007cell, diz2013control}. Actin polymerisation at the membrane gives rise to dynamic structures such as lamellipodia, filopodia and membrane ruffles \cite{blanchoin2014actin, peter2004bar}.
At the edge of adherent cells (in the lamellipodium) actin polymerisation against the membrane causes an actin retrograde flow, which experiences friction with the substrate mediated by adhesion proteins such as integrins, leading to a pushing force on the cell front \cite{lauffenburger:1996,pio:motil_tension}. Remarkably, many cell types such as immune cells can generate thin and extended ($>10\mu$m long) free-standing pseudopods
important for sensing and navigating through complex physiological environments \cite{fritzLaylin2017}. Such actin-filled protrusions, which bear similarities with those observed in reconstituted {\it in-vitro} actin/membrane systems \cite{simon2019actin}, are not connected to any extracellular structures and must rely on purely internal stress within the actin gel in a way that is not understood.

\begin{figure}[b]
    \centering
        \includegraphics[width=\linewidth]{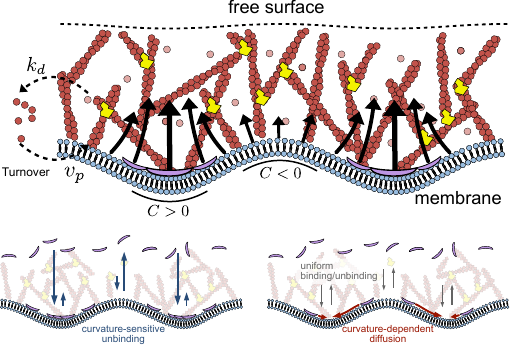}
    \caption[]{Top: Model for a viscous actin gel polymerizing on a wavy membrane. Balance between polymerization  $v_p$  at the membrane and depolymerization $k_d$ yields a layer of finite thickness. Curvature sensitive proteins (purple) locally modulate actin polymerization. Bottom: Two possible scenarios of protein curvature sensing: a curvature-dependent unbinding rate (left) increases the protein bound time in regions of preferred curvature, and/or a spontaneous curvature (right) drives the diffusion of membrane-bound proteins toward regions of preferred curvature.}
    \label{Fig1}
\end{figure}

Active gels growing on surfaces of constant curvature exert a compressive force on the surface whether growing on the convex or concave side of it \cite{sekimoto2004,noireaux2000,julicher2007}. This is a second order effect, hence not sensitive to the sign of the curvature and unlikely to trigger shape instability. Actin retrograde flow in adherent cells can trigger shape instability \cite{callan2008viscous} and even symmetry breaking and waves for non-linear adhesion forces \cite{sens:2020}, but this requires substrate friction. Here we study the stability of a free-standing actin gel growing on a deformable membrane using the theory of active gels \cite{marchetti:2013}. The model, sketched in \cref{Fig1}, includes the modulation of actin dynamics by proteins sensing the local membrane curvature. This is motivated by the fact that many curvature-sensitive proteins such as
BAR-domain proteins can directly or indirectly regulate actin dynamics \cite{zhao:2011,lou2019,tsai2022,sitarska2023}.
Such coupling between actin polymerisation and curvature has been shown to give rise to instability and spontaneous large scale membrane deformation \cite{gov2006,sadhu:2023}, but these models consider that actin polymerisation exerts a net force on the membrane, which is not appropriate for free-standing actin/membrane systems, in which the net force must vanish according to Newton's third law.
Here we show that a gel with uniform polymerisation always flattens the membrane, but that curvature modulation of actin dynamics can render the system linearly unstable, in a way that depends on the interplay between membrane and actin dynamics.

\textbf{Model.} 
Actin is polymerized at the membrane surface which creates a net flow of actin away from the membrane (\cref{Fig1}). 
Actin turnover is included by assuming a constant depolymerisation rate $k_d$ throughout the actin layer and mass conservation is given by $\nabla \cdot(\rho_a \boldsymbol{v}) = - k_d \rho_a$, where $\rho_a$ is the actin density and $\boldsymbol{v}$ the flow velocity. We assume here that the gel is incompressible (see \cite{mihali:long} for the compressible case), such that
\begin{equation}
    \nabla \cdot \boldsymbol{v} = -k_d\;.\label{eq:mass-conservation1}
\end{equation}
The balance between polymerisation and depolymerisation ensures that the actin layer reaches a finite thickness
denoted by $h(x)$ measured as the vertical distance from the membrane to the free surface. 
The growing actin layer is modeled as a viscous fluid described by the linear stress constitutive relation $\boldsymbol{\sigma} = -p\boldsymbol{I}+\eta \left(\nabla\boldsymbol{v} + \nabla\boldsymbol{v}^{\transpose}\right)$,
where $p$ denotes the pressure and $\eta$ the dynamic viscosity.
Here, the pressure acts as a Lagrange multiplier ensuring constant density (cf. \cref{eq:mass-conservation1}).
At low Reynolds numbers and in the absence of external forces, momentum balance is given by $\nabla\cdot \boldsymbol{\sigma} = 0$, which, together with \cref{eq:mass-conservation1} leads to Stokes' equation 
\begin{equation}
    -\nabla p + \eta\nabla ^2 \boldsymbol{v} =0\;.\label{eq:Stokes1}
\end{equation}

We assume that
the layer grows normal to the membrane, and that tangential flow is prohibited by strong friction with the membrane. The case of finite friction shows little difference for an incompressible fluid and is discussed in \cite{mihali:long}. The boundary condition at the membrane, located at $u(x,t)$, is $ \boldsymbol{v}\vert_{z=u(x)} = \left(v_p+\delta v_p(x)+\partial_t u/\sqrt{1+(\partial_x u)^2}\right)\boldsymbol{n}_{m}(x)$,
where $\boldsymbol{n}_{m}(x)$ denotes the unit normal vector at the membrane, $v_p$ denotes a constant polymerization velocity and $\delta v_p(x)$ accounts for small, protein regulated local deviations from $v_p$.
The free surface, located at $W(x) = u(x) + h(x)$, is stress-free (no normal or tangential stress) and its dynamics is given by:
$\boldsymbol{v}\cdot \boldsymbol{n}_{f} = \del_t W/\sqrt{1+(\partial_x W)^2}$. 

In what follows, we assume small membrane deformations, i.e. $|\nabla u|\ll1$ and neglect all terms of $\mathcal{O}(|\nabla u|^2)$ and higher.
The membrane normal is then given by $\boldsymbol{n}_{m} \approx -\partial_x u(x) \boldsymbol{e}_{x} + \boldsymbol{e}_{z}$.
The same strategy can be used for the free surface by replacing $u(x)$ by $W(x)$, as
small deformations of the membrane surface lead to small deformations of the free surface $|\nabla h|\ll1$. Consequently, all quantities may be expanded in independent Fourier modes, whose amplitudes are denoted by the subscript $q$, defined for the membrane shapes as: $u(x) = u_q e^{iqx}$. The validity of the linear regime is discussed in the Supplementary Material (SM).

\begin{figure}[t]
    \centering    
    \includegraphics[width=\linewidth]{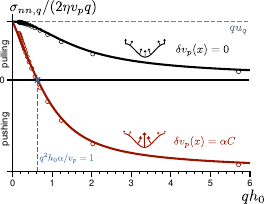}
    \caption[]{Normal stress $\sigma_{nn,q}$ (\cref{eq:total_stress}) exerted by a growing actin layer on a sinusoidal membrane as a function of the dimensionless layer thickness $q h_0=qv_p/k_d$, controlled by varying the depolymerization rate $k_d$. For uniform polymerization (black curve) the stress is positive: pulling at the troughs and pushing at the crests of the wavy membrane, stabilizing the flat shape. 
    If the polymerization velocity varies linearly with the local curvature with a coupling strength $\alpha$ (red curve), the stress changes sign and destabilizes modes satisfying $q^2h_0\alpha /v_p>1$.
    The circles show the numerical results of a finite element method (see SM).
    Parameters: $qu_q = 2\pi/100$, $\tilde{\alpha} = 10/2\pi$}.\label{Fig2}
\end{figure}

\textbf{Steady state actin layer.} The stresses exerted by the polymerizing actin layer on the membrane is obtained by
solving Eqs. (\ref{eq:mass-conservation1},\ref{eq:Stokes1}) 
for the velocity field in terms of Fourier modes (see details in \cite{mihali:long}).
The normal component reads
\begin{equation}
    \sigma_{nn,q}=-2\eta q (k_d u_q +\delta v_{p,q})\tanh\left(qv_p/k_d\right)\;. \label{eq:total_stress}
\end{equation}
An actin layer polymerising uniformly ($\delta v_{p,q} = 0$) on a flat membrane ($u_q = 0$) does not generate any stress ($\sigma_{nn,q} = 0$) and reaches the steady-state thickness $h_0 = v_p/k_d$.
For uniform polymerisation on a wavy membrane shape ($u_q \neq 0$), the stress $\sigma_{nn,q}$ is positive for the entire range of (non-dimensional) layer thickness $q h_0$,indicating that the layer pushes on the hills and pulls on the valleys of the sinusoidal shape, acting to reduce the membrane deformation. The variation of the normalized stress $\sigma_{nn,q}/2 \eta v_p q$ with $qh_0$ is shown in \cref{Fig2}, where we choose to vary the layer thickness by controlling the depolymerization rate $k_d$.
The stress asymptotically reaches zero in the limit of $qh_0\rightarrow \infty$ (corresponding to $k_d \rightarrow 0$),
consistent with previous results \citep{simon2019actin}.
Note that if the layer thickness changes through variation of the polymerization velocity $v_p$, the limiting cases for a thin layer ($v_p$ small or $k_d$ large) and a thick layer ($v_p$ large or $k_d$ small) come with different interpretations (see SM and \cite{mihali:long}). Nevertheless, whichever way the layer thickness is varied, the membrane-actin system under uniform polymerization is linearly stable in the limit of small deformations.

The modulation of actin dynamics by curvature-sensitive proteins may be accounted for at the linear level by introducing a curvature-dependent modulation of the polymerization velocity of the form
$\delta v_{p,q} = -\alpha q^2u_q$ where $\alpha$ is the coupling strength. Inspection of \cref{eq:total_stress} shows that the polymerisation stress changes sign for $q>\sqrt{k_d/\alpha}$. This behaviour is shown in \cref{Fig2} (blue line). The numerical solution of this system of equations using a finite element method (detailed in the SM) confirms these results and show excellent agreement with the linearized analytical solutions (\cref{Fig2}).

\textbf{Coupling actin and membrane dynamics.} To understand how this mechanism can drive spontaneous membrane deformation, we derive the membrane shape dynamics by comparing the actin-generated forces to the membrane restoring forces \cite{deserno2015} including the dynamics of curvature-sensitive proteins modulating actin dynamics. We consider a mixture of positively curved and negatively curved proteins, denoted by $\rho_{+}$ and $\rho_{-}$, with spontaneous curvature $\pm H$, respectively. We assume that the average protein coverage is the same for both protein types ($\int dS(\rho_+-\rho_-)=0$), so that the membrane under uniform protein coverage is flat. In this case, the protein density field relevant to membrane deformation only involves the signed protein density $\rho=\rho_+-\rho_-$ in the limit of small deformation (see \cite{ramaswamy2000nonequilibrium} and SM for details). The membrane free energy includes an elastic contribution based on Helfrich free energy, a protein contribution written as a Ginzburg-Landau expansion and a coupling term due to protein spontaneous curvature
\cite{helfrich1973, marcerou1983,leibler1986curvature,andelman1992}
\begin{align}
    \mathcal{F}[\rho, u] = \int d^2x \Bigl[& \frac{\kappa}{2}(\nabla^2u)^2 +\frac{\gamma}{2}(\nabla u)^2\notag\\ &-\kappa \rho H \nabla ^2 u + \frac{a}{2}\rho^2 + \frac{b}{2}(\nabla \rho)^2\Bigr]\label{eq:Helfrich_MP} \;,
\end{align}
where $\kappa$ and $\gamma$ are the membrane bending rigidity and tension, and $a$ and $b$ are the strength of protein-protein interaction and cost of density gradients. 
From \cref{eq:Helfrich_MP} we obtain the membrane restoring stress $\sigma_{m} = -\delta \mathcal{F}/\delta u = -\kappa \nabla^4 u + \gamma \nabla^2 u + \kappa H \nabla^2 \rho$ and the generalised protein chemical potential $ \mu = \delta \mathcal{F}/\delta \rho =-\kappa H\nabla^2 u + a \rho - b\nabla^2\rho$. The ratio $\sqrt{b/a}$ defines a length scale typically of order the protein size. As we are concerned with concentration modulation at (much) larger length scales, we neglect the concentration gradient energy term ($b=0$).

We assume a linear relationship between actin polymerization modulation and the signed protein density with a coupling strength $A$:
\begin{equation}
    \delta v_p = A \rho\;.
\end{equation} 
Demanding force balance between actin and membrane stresses: $\sigma_{m} + \sigma_{nn} = 0$,
we obtain a dynamical equation for the membrane deformation.
The time evolution of the signed membrane proteins density is governed by mass conservation $\del_t\rho = \Lambda\Delta\mu + j_r$, where $ j_r$ is the recycling flux for the signed density, which may depend on curvature (see \cref{Fig1}). The recycling flux for each protein type is written as $j_{r,\pm}=j_0/2-k_{\rm off}^0(1\mp \lambda_{\rm off} C)\rho_\pm$, such that proteins with positive spontaneous curvature unbind slower from regions of positive curvature \cite{jin2022}, with a coupling strength $\lambda_{\rm off}$. On a flat membrane, the average steady-state total protein density is $\rho_0=j_0/k_{\rm off}^0$ and the recycling flux for the signed density is at lowest order $j_r=k_{\rm off}^0\rho_0(-\rho/\rho_0+\lambda_{\rm off} C)$.

The equations are made dimensionless using the membrane mechanical length scale
$\lambda = \sqrt{\kappa/\gamma}$ \cite{pio:motil_tension} as the characteristic length scale, $\lambda/v_p$ as the time scale and $2 \eta v_p/\lambda$ as the stress scale. All resulting variables are expressed in dimensionless form and are denoted by a bar (e.q. $\bar{u}$, $\bar{\sigma}$, etc., see table S1). The dynamics of the system in Fourier space reads
\begin{widetext}
\begin{equation}
\del_{\bar t}\begin{pmatrix}
\bar{u}_q \\
\rho_q
\end{pmatrix}
=
\begin{pmatrix}
- \frac{1}{\bar{h}_0} - \frac{\bar{\gamma}(\bar{q} + \bar{q}^3) }{\tanh(\bar{q} \bar{h_0})} & -\frac{ \bar{\gamma} \bar{H} \bar{q} }{\tanh(\bar{q} \bar{h_0})} - \bar{A}\\
-\bar{\Lambda} \bar{\gamma} \bar{ H} \bar{q}^4 -\bar{k}_{\text{off}}^{\text{0}}\bar\lambda_{\text{off}} \rho_0 \bar{q}^2& -\bar{\Lambda}\bar a\bar q^2-\bar{k}_{\text{off}}^{\text{0}} \\
\end{pmatrix}
\begin{pmatrix}
\bar{u}_q \\
\rho_q \label{eq:dynamical_matrix}
\end{pmatrix}
\;.
\end{equation}
\end{widetext}
As the trace of the dynamical matrix in \cref{eq:dynamical_matrix} is always negative, the fixed point of the dynamical system is stable if the determinant of the matrix is positive, and turns into an unstable saddle point if $\Det < 0$.

\begin{figure}[t]
    \centering
    \includegraphics[width=\linewidth]{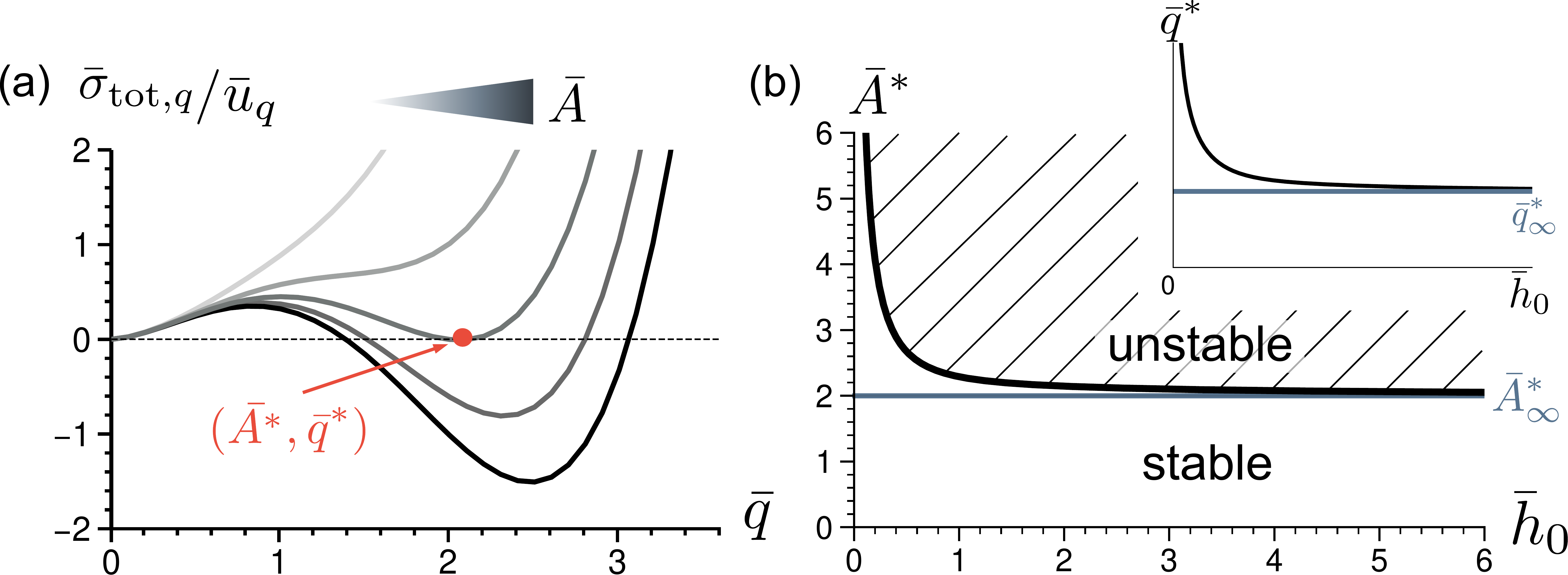}
    \caption[]{
    Total steady-state membrane stress for fast protein diffusion without recycling.
    (a)  Total normal stress as a function of the wave number $\bar{q}$ for different coupling strengths $\bar{A}$.
    A negative value indicates a destabilizing effect.
    (b) Critical coupling strength $\bar{A}^{\ast}$ at which the first unstable mode appears (wavevector $q^*$, red dot in panel (a)) as a function of the layer thickness. Inset shows the value of $q^*$.
    Parameters: $\bar{H}=1$, $\bar{a}=2$,
    $\bar{\gamma}=1.5$ and $\bar{h}_0 = 10$ for panel (a).}
    \label{Fig3}
\end{figure}

We first analyze the case of slow protein recycling and fast protein diffusion ($\Lambda\rightarrow\infty$, $k_\text{off}^0\rightarrow 0$). Driven by the spontaneous curvature $H$, the protein density quickly adjusts to its steady-state distribution proportional to the local membrane curvature for any membrane shape: $\rho_q = -\bar{q}^2 \bar{u}_q \bar{\gamma}\bar{H}/\bar{a}$. Even in the absence of protein/actin coupling ($A=0$), this phenomenon is known to reduce the effective membrane bending rigidity, and can lead to the so-called ``curvature instability'' beyond a spontaneous curvature threshold, when the effective rigidity becomes negative and large membrane deformation spontaneously occurs \cite{leibler1986curvature}. Here we concentrate on the role of actin and remain below this threshold ($\bar H<\sqrt{\bar a/\bar\gamma}$ \cite{leibler1986curvature}).
The total steady-state membrane stress is shown in \cref{Fig3} (a) as a function of the wavenumber for different coupling strengths $\bar{A}$.
For weak coupling ($\bar A\ll1$), both the actin stress and the membrane stresses stabilize the flat membrane, so the stress remains of a given sign (positive - pulling - on the valleys of a sinusoidal shape). For strong enough coupling, the actin stress can destabilize a finite range of $q$ modes. Since the membrane stress stabilizes large $q$ modes, there exists a critical coupling strength $\bar{A}^{\ast}$ beyond which a range of modes around a critical wave mode $q^*$ become unstable. This value is shown as a function of the layer thickness in \cref{Fig3} (b), together with the corresponding value of $q^*$. Both are decreasing functions of the gel thickness, with asymptotic values for $q^*\bar{h}_{0} \gg1$
\begin{equation}
\bar{A}^{\ast}_{\infty} = 2\frac{\bar a}{\bar H}\sqrt{1-\bar H^2\bar\gamma/\bar a}\ ,\ \ \bar{q}^{\ast}_{\infty} = 1/\sqrt{1 - \bar{H}^2 \bar{\gamma}/\bar{a}}\;.
\label{eq:Acrit_steady_state}
\end{equation} 
We see that $A_\infty^*\rightarrow0$ as $\bar H\rightarrow\sqrt{\bar a/\bar\gamma}$, consistent with the classical curvature instability discussed above \cite{leibler1986curvature}.

In the opposite limit of very slow protein diffusion ($\Lambda\rightarrow 0$, $k_\text{off}^0\rightarrow\infty$), the instability must rely only on curvature-dependent recycling, and one may show (see SM) that the instability threshold (for thick layers) is given by
\begin{equation}
\bar{A}^{\ast}_{0} = \frac{2\bar\gamma}{\rho_0\bar{\lambda}_{\rm off}}\sqrt{1- \rho_0 \bar H\bar{\lambda}_{\rm off}},\ \ \bar{q}^{\ast}_{0} = 1/\sqrt{1 - \rho_0\bar H\bar{\lambda}_{\rm off}}.
\label{eq:Acrit_recycling}
\end{equation} 
This expression indicates a new type of actin-independent curvature instability, mediated by the curvature-dependent turnover of curvature-active proteins, which occurs if $\rho_0 H\lambda_\text{off}>1$ (see SM). Here again, we remain below this limit and concentrate on the role of actin.

\begin{figure}[h]
    \centering
    \includegraphics[width=\linewidth]{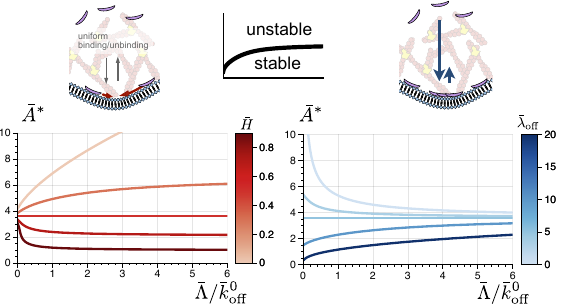}
    \caption[]{ Instabilities threshold $A^*$ (the system is unstable if $A>A^*$) as a function of the ratio of protein mobility to recycling rate $\bar\Lambda/\bar k_{\rm off}^0$. One the left panel, the spontaneous curvature $\bar H$ is varied.
    at fixed curvature-sensitive unbinding rate $\bar \lambda_{\rm off}=5$. On the right panel, $\bar \lambda_{\rm off}$
    is varied at constant $\bar H =0.5$. The asymptotic regimes of low and high mobility correspond to Eqs. (\ref{eq:Acrit_steady_state},\ref{eq:Acrit_recycling}). The instability is eased by protein mobility if $\bar H>\bar a\rho_0\bar \lambda_{\rm off}/\bar\gamma$ (see text).
    Parameters: $\bar{a}=1$, $\bar{b}=0$, $\bar{\gamma}=1$, $\rho_0=0.1$, $\bar{h}_0=10$.}
    \label{Fig4}
\end{figure}

The coupling threshold $A^*$ is shown as a function of the ratio of in-plane diffusion to recycling rates $\bar\Lambda/\bar k_{\rm off}^0$ in \cref{Fig4}, for the two curvature-sensing mechanisms displayed in \cref{Fig1}. It varies monotonously between the asymptotes given by Eqs.~(\ref{eq:Acrit_steady_state},\ref{eq:Acrit_recycling}). Therefore the instability is promoted by fast protein diffusion if protein localisation by membrane curvature is more efficient via in-plane diffusion than via turnover, which corresponds to $\bar A^*_\infty<\bar A^*_0$, or $\bar H>\bar a\rho_0\bar \lambda_{\rm off}/\bar\gamma$.
The left panel of \cref{Fig4} shows this behavior varying $\bar H$ at fixed $\bar\lambda_{\rm off}$ and the right panel varying $\bar \lambda_{\rm off}$ at fixed $\bar H$. As expected, the system is always stable ($A^*\rightarrow\infty$) for very slow diffusion in the absence of curvature-dependent recycling ($\bar \lambda_{\rm off}=0$) and for very fast diffusion without spontaneous curvature ($\bar H=0$).

\textbf{Discussion.}
Our results may be summarized as follows: an actin gel growing on a corrugated membrane exerts a mechanical stress on the membrane. For an incompressible gel growing uniformly, the stress stabilizes flat membranes and can reach levels of order the elastic modulus of the gel $E\sim\eta k_d\simeq 10^4$Pa for a gel of viscosity $\eta=10^4$Pa.s and of turnover rate $k_d=1 {\rm s}^{-1}$ \cite{simon2019actin}. Spontaneous actin-driven protrusion (linearly unstable) are predicted to occur in case actin dynamics (polymerisation rate) is sensitive to the curvature of the membrane it grows from. 
Curvature-sensing proteins are known to modulate actin polymerisation \cite{sitarska2023, lou2019}. Both N-BAR (Amphiphysin) and I-BAR (IRSp53, MTSS-1), respectively attracted to negative and positive curvatures according to our convention, can promote actin recruitment \cite{sorre2012, prevost2015}. Curvature sensing of BAR domains involve both a spontaneous curvature and curvature-dependent recycling. We account for both effects (through the parameters $\bar H$ and $\bar \lambda_{\rm off}$, respectively) 
within a full linear model coupling actin and protein dynamics with membrane mechanics.

Our model shows the existence of an actin-driven curvature instability when proteins sensitive to positive curvature promote polymerisation. This is in line with observations that I-BAR is implicated in protrusion formation \cite{zhao:2011,tsai2022}. This instability is predicted to occur if the value of the parameter $\bar A$, linearly coupling actin polymerisation modulation to the curvature-dependent protein density (surface fraction) $\rho$ according to $\delta v_p/v_p=\bar A \rho$, is beyond a threshold value $A^*$. Asymptotic expressions when either of the two curvature-sensing mechanisms dominates are given in \cref{eq:Acrit_steady_state,eq:Acrit_recycling}, and the full phase diagram is shown \cref{Fig4}.
Which mechanism dominates is largely controlled by the ratio of protein mobility to recycling rate.
As the latter varies over a wide range for BAR proteins ($k_{\rm off}^0:\ 0.01-10/$sec \cite{jin2022}, corresponding to $\bar \Lambda/\bar k_{\rm off}^0:\ 0.1-100$), this shows the relevance of exploring this parameter as done in \cref{Fig4}.
For appropriate physical parameters we find that $\bar A^*\simeq 5-10$ which corresponds to a doubling of actin polymerisation speed in regions of favored curvature (with $\rho\simeq 10\% $). This seems like a reasonable threshold, reachable in cellular systems. The typical length scale of the first unstable mode is of order $2\pi/q^*\simeq 2\pi\lambda\simeq 500$nm. It is comparable to the typical thickness of free-standing cellular protrusions (thickness of $430-800\,\text{nm}$ reported in \cite{fritzLaylin2017}). Although the latter certainly result from non-linear coarsening effects beyond the linear stability analysis proposed here, such comparison is rather encouraging.

\textbf{Acknowledgements.}
This project has received funding from the European Research Council (ERC) 
ERC-SyG (Grant agreement ID: 101071793)

\providecommand{\noopsort}[1]{}\providecommand{\singleletter}[1]{#1}%
%



\twocolumngrid

\end{document}